\documentclass[pre,twocolumn,superscriptaddress,noshowpacs]{revtex4-2}
\usepackage{graphicx}
\usepackage{physics}
\usepackage{bm}
\usepackage[utf8]{inputenc}

\begin{document}
\title{Love might be a second-order phase transition}

\author{D.~D.~Solnyshkov}
\affiliation{Institut Pascal, PHOTON-N2, Universit\'e Clermont Auvergne, CNRS, Clermont INP, F-63000 Clermont-Ferrand, France.}
\affiliation{Institut Universitaire de France (IUF), F-75231 Paris, France}

\author{G.~Malpuech}
\affiliation{Institut Pascal, PHOTON-N2, Universit\'e Clermont Auvergne, CNRS, Clermont INP, F-63000 Clermont-Ferrand, France.}

\begin{abstract}
The hypothesis of the human brain operation in vicinity of a critical point has been a matter of a hot debate in the recent years. The evidence for a possibility of a naturally occurring phase transition across this critical point was missing so far. Here we show that love might be an example of such second-order phase transition. This hypothesis allows to describe both love at first sight and love from liking or friendship.
\end{abstract}
\maketitle

\section{Introduction}

It is natural to try to use the well-developed methods of statistical physics when one is dealing with large numbers of entities exhibiting relatively simple interactions \cite{kwapien2012physical}: financial markets, natural languages, and neural networks. 
Such methods have been applied to artificial and natural neural networks almost since the very beginning of their studies. Indeed, a natural neural network represented by the human brain contains $10^{11}$~neurons, which, while being smaller than the number of atoms in a typical macroscopic object $N_A=6\times 10^{23}$, still falls well into the domain of applicability of statistical physics \cite{Kinzel}. Quite soon after the beginning of the studies of artificial neural networks, such as perceptrons, it was shown that the learning process of such networks was mathematically equivalent to a Monte-Carlo procedure to obtain thermal equilibrium \cite{Hopfield1982,amit1987statistical}.

The description of larger neural networks, such as the whole human brain or its major elements, such as the cortex, advanced thanks to experimental measurements of the electric activity of the brain. These measurements, as well as the studies of externally accessible behaviors, were showing the so-called $1/f$ noise \cite{gilden19951}. This power law has become one of the major indications of the criticality of the brain's behavior \cite{kello2007emergent}. Indeed, one of the leading hypotheses now is that the brain is operating close to the critical point of a second-order phase transition \cite{kelso1992phase,PhysRevE.79.061922,chialvo2010emergent,beggs2012being,cocchi2017criticality}, which explains several groups of its properties: the presence of power-law scalings for many of its characteristics \cite{gilden19951}; the links between the corresponding scaling exponents, including those describing behavior \cite{palva2013neuronal}; the emergence of large-scale dynamical correlations \cite{linkenkaer2001long,PhysRevLett.110.178101} between neurons (pattern formation) and the sensitivity to the state of individual neurons; the divergence of the susceptibility and the multitude of self-adapting reactions. The operation at the critical point allows the brain (and other biological systems) to achieve maximal efficiency, required for survival \cite{moretti2013griffiths}. It is ensured by the proper balance of excitation and inhibition \cite{PhysRevLett.96.028107,poil2012critical}

On the other hand, the criticality is not universally accepted yet. One of the alternatives to the criticality is the self-organized bistability \cite{PhysRevResearch.2.013318}. Power laws, and in particular, the $1/f$ scaling, can emerge in networks even without criticality \cite{PhysRevLett.97.118102,PhysRevE.95.012413}. Yet another argument against the criticality theory is that it implies the possibility to cross the transition point, and thus one can expect some evidence of such transitions. Ongoing studies are devoted to the type of the phase transition involved \cite{PhysRevLett.122.208101,PhysRevX.11.021059,PhysRevE.100.032414}.  It is understood that the balance between excitation and inhibition should be one of the parameters allowing to cross the transition point.  An artificially-induced transition to the subcritical regime and a return to criticality have already been observed in rats \cite{ma2019cortical}, but no evidence of such transitions occurring naturally are known for human beings so far. 

In this work, we suggest that love might be an example of a second-order phase transition occurring in the brain. We show that this hypothesis explains a lot of well-known properties of love. Analyzing several most famous literature examples and a private diary, we show that the intensity of feelings exhibits a universal scaling behavior, distinguishing two cases: love at first sight and love developing from liking or friendship (friends first), both being studied in psychology \cite{Barelds2007,Hefner2013,Vannier2017}.

\section{Second-order phase transitions}

\begin{figure}
    \centering
    \includegraphics[width=0.99\linewidth]{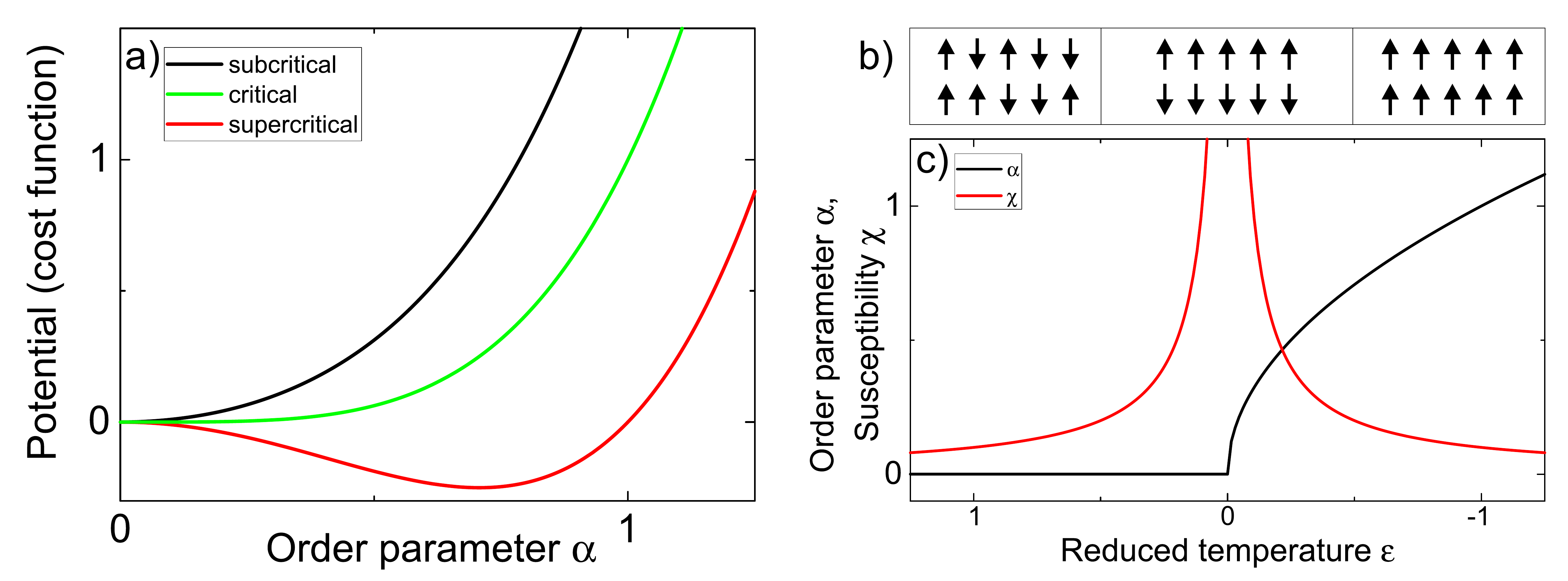}
    \caption{Second-order phase transitions. a) Generalized potential $F$ as a function of order parameter $\alpha$. The minimum is at $\alpha=0$ "above" transition (subcritical behavior) and at $\alpha\neq 0$ "below" transition (supercritical behavior). b) Ising model of a ferromagnet illustrating the 3 regimes (subcritical/critical/supercritical). c) The order parameter $\alpha$ and the susceptibility $\chi$ as a function of a dimensionsless parameter (reduced temperature $\epsilon$).}
    \label{fig1}
\end{figure}

The theory of second-order phase transitions developed by Landau \cite{Landau5} states that the thermodynamic potential $F$ (that we generalize to an abstract "cost function" minimized by the system) can be written in series expansion in powers of the order parameter $\alpha$:
\begin{equation}
    F\left(\alpha\right)=F_0+h\alpha+A\alpha^2+B\alpha^4+\ldots
    \label{pot}
\end{equation}
In general, $h$ corresponds to an external bias (e.g. applied field) that we will discuss later. A phase transition of the second order (or kind) occurs when the coefficient $A$ (depending on a certain parameter, usually temperature) changes sign, crossing zero. This is illustrated in Fig.~\ref{fig1}(a), showing the typical shapes of $F$ for $A>0$ (black), $A=0$ (green), $A<0$ (red) for $h=0$. Since $A$ changes sign, it necessarily behaves as a first power of temperature (or reduced temperature $\epsilon=T/T_c$, where $T_c$ is the critical temperature, where the transition occurs): $A\sim \epsilon$. The minimum of $F$ is determined by $\partial F/\partial \alpha=0$ giving a cubic equation
\begin{equation}
    \alpha^3+\frac{A}{2B}\alpha+\frac{h}{4B}=0
    \label{cub}
\end{equation}
For $h=0$, the minimum is located at $\alpha=0$ for $A\ge 0$ and at $\alpha\neq 0$ for $A<0$. Spontaneous symmetry breaking occurs at $A=0$.

This theory has inspired a lot of activity in physics, because it turned out that such transitions are characterized by a universal scaling behavior. Independently of the nature of the system, their parameters exhibit power law dependencies on the dimensionless parameter (temperature). For example, the order parameter $\alpha$ scales as a square root of temperature (below threshold): $\alpha\sim\sqrt{|\epsilon|}$. Another important parameter, the susceptibility, diverges at the critical temperature as $\chi\sim|\epsilon|^{-1}$. It means that the order parameter $\alpha$ reacts strongly to any external perturbation (introduced via the term $h$). This is one of the reasons why operating close to the critical point $T_c$ ($\epsilon=0$) is thought to increase the efficiency of the brain.

The second-order transitions are best illustrated by the Ising model of a ferromagnet, consisting of interacting spins (Fig.~\ref{fig1}(b)). The fact that a neural network can be mapped to this model was discovered quite a long time ago~\cite{Little1974}. The transition to the ordered phase usually occurs while decreasing the temperature.
Above the critical temperature, the minimum of the generalized potential $F$ corresponds to zero order parameter $\alpha=0$, and thus the system exhibits no order (random orientation of different spins). Because of this, the system is not sensitive to the orientation of a single spin. This is the subcritical regime. Below the critical temperature, the order parameter (average magnetization) is non-zero $\alpha\neq 0$: all spins are pointing in the same direction (chosen randomly at the moment of spontaneous symmetry breaking). The sensitivity to a single spin is low: all spins are aligned, and a single spin flip does not perturb them all. This is the supercritical regime. Finally, exactly at the threshold the system exhibits long-range patterns and a high sensitivity to a single spin: flipping it starts flipping all other spins in the pattern. This is the critical regime. It is also characterized by divergent fluctuations of the order parameter $\langle \Delta \alpha^2\rangle\sim |\epsilon|^{-1}$ (due to their low energy cost). For neural networks, where each spin is mapped to a neuron, this implies high sensitivity to a state of an individual neuron combined with long-range pattern formation, which corresponds to a high memory capacity and high adaptability of the system.

If one wishes to study the second-order phase transition process, and not the individual phases above and below it, the reduced temperature $\epsilon$ has to vary with time, which becomes the new dimensionless parameter: $A\sim -t/\tau$ (where $\tau$ is the characteristic transition time also called quench time). The variation of reduced temperature should be sufficiently slow, so that a finite-size system can reach equilibrium at each moment (we neglect the divergence of the relaxation time, known to play a strong role in the Kibble-Zurek mechanism of topological defect formation in the ordered phase \cite{Kibble1976,Zurek1985}). Studying a time-dependent system, one can therefore expect the order parameter to be zero until the transition, to exhibit large fluctuations at the transition, and then to grow as a square root of time $\alpha\sim \sqrt{t}$.

\section{Love at first sight}

The key hypothesis of the present work is that love is a second-order phase transition occurring in the human brain under the influence of hormones, such as dopamine and serotonin  \cite{takahashi2015imaging,de2012love}. The brain switches from the normal operation in the critical regime to the supercritical regime because of the increase of the excitation (more dopamine) and reduction of the inhibition (less serotonin) \cite{de2012love}. The order parameter of the transition is the intensity of feelings: before the transition, the subject has no particular feelings, while after the transition the feelings are non-zero.
The most direct consequence is that the order parameter should exhibit a universal power law behavior: the feelings should grow as a square root of time 
\begin{equation}
    \alpha\sim \sqrt{t}
\end{equation}
for all individuals, in the absence of an external bias ($h=0$).

Here we spot the first indirect confirmation of our hypothesis. Indeed, a square root time dependence $\alpha\sim\sqrt{t}$ is characterized by an infinitely fast initial growth: $\partial \alpha/\partial t\sim 1/\sqrt{t}$, giving $\partial \alpha/\partial t=\infty$ at 
$t=0$. This infinitely fast growth of feelings is striking for the individual who feels it. This striking feeling is probably at the origin of the expression "love at first sight". It is also why the Roman god Cupid (or Amor), called Eros in Greek, carries a bow with arrows. The strike of Cupid's arrow represents the fast growth of feelings suggested by the model. Finally, in French "love at first sight" is translated as "le coup de foudre" (the strike of a lightning), confirming the hypothesis further. According to the prediction of psychological theories , "love at first sight" should be marked with a particularly high passion \cite{zsok2017kind}, which agrees with this hypothesis.

Since it is impossible to analyze the human feelings directly during such uncontrollable phenomenon as "love at first sight", we have decided to perform a quantitative study of the intensity of feelings depicted in the literature. This indirect approach is similar to the one used to detect the $1/f$ noise via measurement errors and reaction time in the seminal paper of D.~L.~Gilden \cite{gilden19951}. We have chosen three well-known books for this analysis: "Romeo and Juliet" (W. Shakespeare), "The Lily of the Valley" (H. de Balzac, see Appendix~\ref{appLily}), and "Martin Eden" (J. London).
While all three books are fiction, we assume that the descriptions of feelings of main characters should be true enough, otherwise they would seem "wrong" to the reader and would not gain such popularity. Moreover, H. de Balzac is known for the knowledge of the human nature and for the attention to detail, whereas the book of J. London is considered as autobiographic and should also be true in its description of love.

From each of the books, we have chosen several consecutive phrases from which a reader can deduce the feelings of the main character, and submitted these phrases to a set of six to ten (different for different texts) female and male respondents aged from 40 to 83 years (that is, used to reading texts instead of watching movies and YouTube) and unaware of the theoretical hypothesis being tested, in order to avoid subconscious or conscious bias. The respondents were asked to assess the intensity of feelings of the main character for each phrase using a scale from 0 to 10. We have normalized the maximal intensity of feelings to $1$ for each book (the scaling only holds close to the transition point and cannot describe the evolution of feelings for longer times), and we assume that this intensity is achieved at $t=\tau$. We assumed a constant time spacing between consecutive phrases: we suppose that while the speed of perception can be different, it cannot change over such a short period, and the speed of the narrative corresponds to the natural perception. 

\begin{figure}
    \centering
    \includegraphics[width=0.99\linewidth]{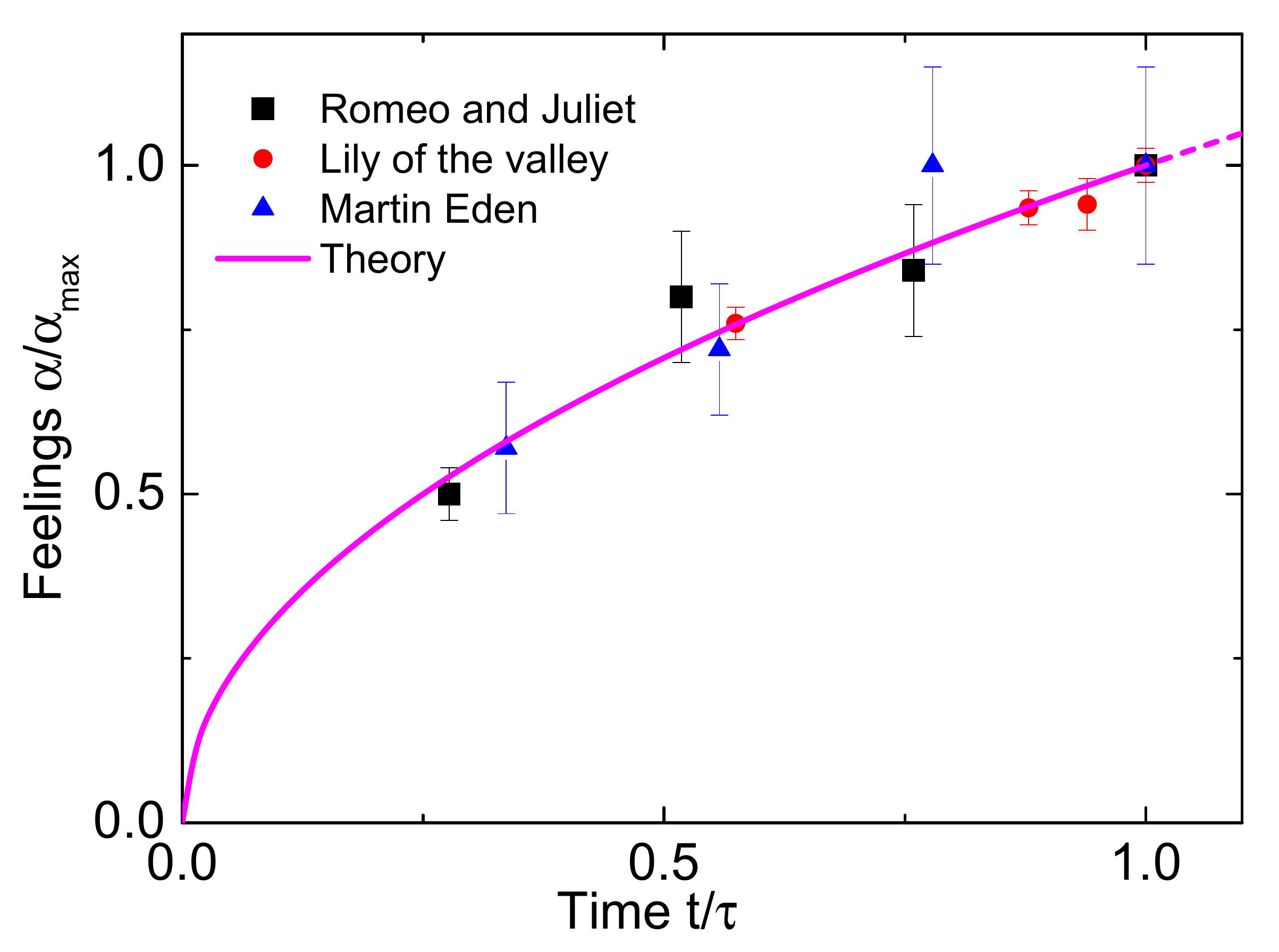}
    \caption{The intensity of feelings demonstrating universal scaling exponent $1/2$ as a function of the variable parameter (time) for 3 books (black -- "Romeo and Juliet", red -- "The Lily of the Valley", blue -- "Martin Eden"). Magenta curve -- theory. The error bars indicate the confidence intervals.}
    \label{fig2}
\end{figure}

The fitting parameter for each book was the moment at which the transition occurs, because it is not mentioned explicitly in these books. The results are shown in Fig.~\ref{fig2}. The 3 sets of points correspond to the 3 books, and the magenta curve to the theoretical scaling $\alpha/\alpha_0=\sqrt{t/\tau}$. The single theoretical curve showing  the predictions of the second-order phase transition model is universally compatible with all 3 books. The error bars show the confidence intervals at 0.95~level.

Let us discuss these examples in more details, so that the reader can become convinced that the described feelings indeed grow as assessed by our respondents. Indeed, in Romeo and Juliet one finds the following phrases:
\begin{enumerate}
    \item What lady's that, which doth enrich the hand of yonder knight?
    \item O, she doth teach the torches to burn bright!
    \item  she hangs upon the cheek of night
    Like a rich jewel in an Ethiop's ear, beauty too rich for use
    \item for earth too dear
\end{enumerate}
The continuous growth of feelings can be easily noticed in this sequence, from the first phrase, merely noticing that the lady "enriches" the hand of a knight, to the last phrase, where Juliet is compared to a jewel "too dear for Earth" (which can be naturally taken as a maximum). Especially important is the fast initial increase illustrated by the second phrase.

It is interesting to note that the Romeo's behavior  exhibits signatures of large fluctuations $\langle \Delta \alpha^2\rangle\sim |\epsilon|^{-1}$ close to the transition point: before falling in love with Juliet, he loves others, although not to the same extent (not because they were worse than Juliet, but because he has not yet crossed the transition point). These short episodes also agree with the predictions of the theory.

From "Martin Eden", which has the advantage of being autobiographic,  we have selected the following phrases:
\begin{enumerate}
    \item He did not know how
she was dressed, except that the dress was as wonderful as she
    \item He
likened her to a pale gold flower upon a slender stem.
    \item No, she was a
spirit, a divinity, a goddess
    \item such sublimated beauty was not of the
earth
\end{enumerate}
Again, a fast growth with saturation is clearly visible. The last phrase clearly expresses the maximal feelings possible ("not of the Earth", as in "Romeo and Juliet"), justifying taking it as $\alpha_{max}$. It is also interesting to note that Jack London himself was interested in scientific analysis of the nature of love, which he discussed in the "Kempton-Wace Letters" \cite{london1991kempton}. We can therefore expect his description of his own feelings to be quite precise. Moreover, in this book we even find a phrase, which indirectly confirms our hypothesis of second-order transition from criticality to super-criticality: "The phantasmagoria of his brain vanished at sight of her".

We note that fitting the whole set of points with a power law in a log-log scale gives a scaling exponent of $0.56\pm0.06$, compatible with the theoretical prediction of $1/2$, which  further confirms the hypothesis of the second-order phase transition. Finally, it is interesting to note that the red points, corresponding to the "Lily of the valley" by H. de Balzac fall almost exactly on the theoretical curve. If the theoretical hypothesis is correct, it confirms the reputation of H. de Balzac as a connoisseur of the human heart. 

The hypothesis of second-order phase transition provides an explanation for other well-known features that characterize love, such as "love is blind". Indeed, the loss of sensitivity associated with the transition to super-criticality essentially makes the person "blind" to the defects of the beloved one, and also "deaf" to the warnings of parents and friends. This is also confirmed by recent psychological studies, demonstrating that falling in love typically involves cognitive biases \cite{barelds2011assessment,fletcher2010through}. 
Other examples of the proverbs explained by the hypothesis include "There is no difference between a wise man and a fool when they fall in love", and "When love is not madness, it is not love." Most probably, contrary to the other aspects, for which the brain is operating in the critical regime, in the case of love the evolution has, on the contrary, avoided the stabilization at the critical point. Indeed, for evolution, the desire for procreation should overcome reason, as well as all other instincts, including the survival instinct. Before maturity, the brain is probably functioning subcritically in what concerns love, and when the maturity is reached, it goes supercritical.
 
We note that in the case of "love at first sight", the transition occurs inside the person and is caused by internal reasons. An individual is moving gradually across the transition with age. We see new faces all the time, but when the person becomes "ready" and crosses the transition point, the particular new face seen at this particular moment becomes the object of spontaneous symmetry breaking, which means that this effect is essentially random. The person could have seen somebody else (not completely disgusting) at this moment  and  would fallen in love in that other person. This is confirmed by the saying "Beauty is in the eye of the beholder". Juliet seemed exceptional to Romeo not because she was exceptional, but because he has just crossed the transition point inside himself, that is, he has moved to an exceptional state himself.

\section{Love from liking}

\begin{figure}
    \centering
    \includegraphics[width=0.99\linewidth]{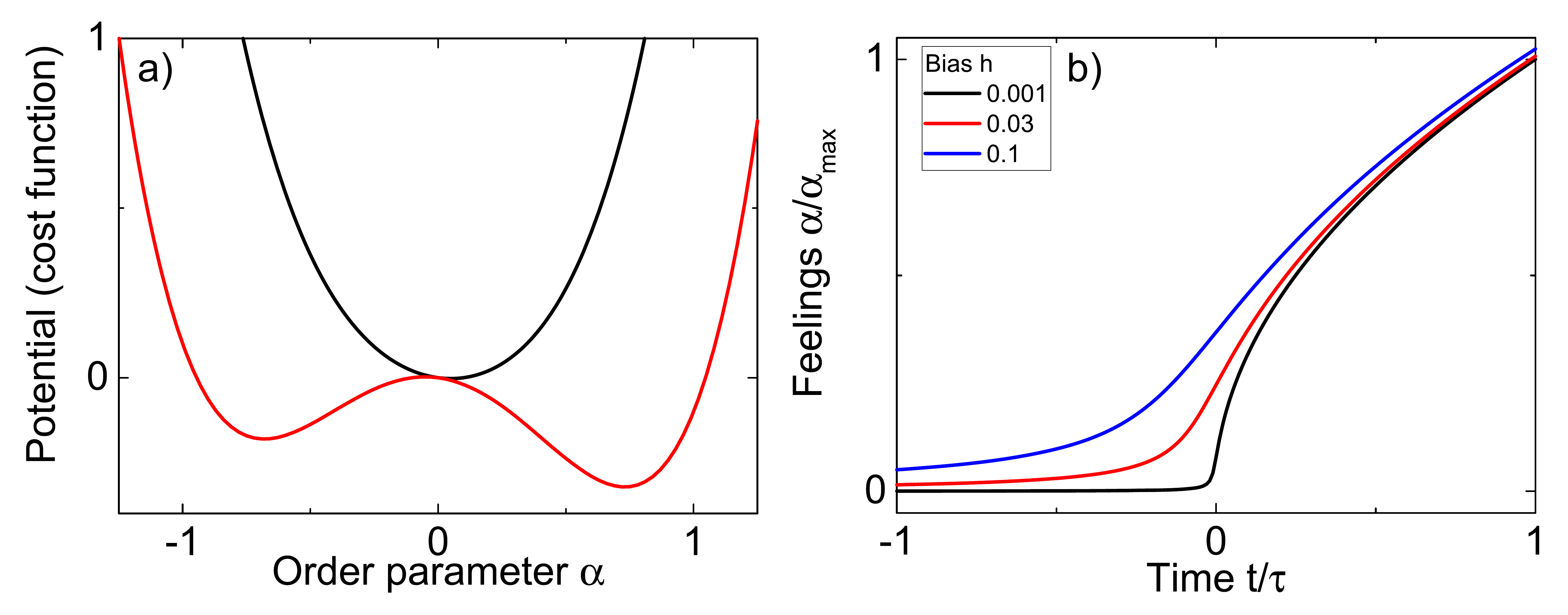}
    \caption{Love from liking. a) The generalized potential $F$ in presence of an external bias $h$: the minimum is at $\alpha\neq 0$ both "above" and "below" threshold. Note the presence of a second (negative) minimum "below" threshold. b) The evolution of feelings $\alpha(t)$ across the transition for different values of $h$ indicated in the figure legend.}
    \label{fig3}
\end{figure}

We now consider the case of a non-zero external "bias", which is a situation that occurs more often than love at first sight. Indeed, one usually knows several persons better than all the others, and thus some positive feelings towards some of them can be present before the transition occurs. The generalized potential $F(\alpha)$ for non-zero $h$ is plotted in Fig.~\ref{fig3}(a). The minimum of $F$ is now always at non-zero $\alpha$. However, when the transition is crossed, this value of $\alpha$ rapidly increases. In general, non-zero $h$  "broadens" the transition shown in Fig.~\ref{fig1}(c). The main root of \eqref{cub} reads
\begin{equation}
    \alpha=\frac{-2\cdot 3^{2/3}A+3^{1/3}\left(-9\sqrt{B}h+\sqrt{24A^3+81Bh^2}\right)^{2/3}}{6\sqrt{B}\left(-9\sqrt{B}h+\sqrt{24A^3+81Bh^2}\right)^{1/3}}
\end{equation}
As previously, the parameter $A$ crosses zero and exhibits a linear dependence on time: $-A\sim t$. The behavior of this solution $\alpha(t)$ is shown in Fig.~\ref{fig3}(b) for different values of $h$.  On the one hand, in this case the transition is, in general, slower, and it is more difficult for the individual to distinguish the moment when liking or friendship turns into love. On the other hand, the exact moment of the transition is characterized by the maximal derivative (infinite in the case of $h=0$). This rapid increase of feelings can hardly go unnoticed, and it is the moment when one understands that love has come.

To study this transition quantitatively, we have chosen another well-known book, "Jane Eyre" by Charlotte Bront\"e. This books is also quite autobiographic, and famous for the psychological details, that got its author the title of "first historian of the private consciousness". The delay between the initial acquaintance and love is now of several weeks (about 45 days). The bias  is clearly present $h\neq 0$, because Mr. Rochester is the only male person around Jane Eyre. We have selected several paragraphs from several chapters narrating the process of falling in love. The advantage of the detailed narrative is that it is possible to date these paragraphs quite precisely (see Appendix~\ref{appendixJE} for the phrases and dates).

The progressive change in the attitude towards Mr. Rochester is clearly visible in these paragraphs. One can also notice the strong change that occurs around Day 28, which we therefore choose as the origin of the transition $t=0$. As before, we normalize the intensity to the maximum, and the time scale to $\tau=45-28=17$~days (the duration between the transition and the maximal intensity). The points are presented in Fig.~\ref{fig3}(b) together with the theoretical curve. The model has a single fitting parameter $h$, for which we find $h\approx -0.68\pm 0.08$ with a p-value $4\times 10^{-5}$, clearly allowing to reject the null hypothesis. 
\begin{figure}
    \centering
    \includegraphics[width=0.99\linewidth]{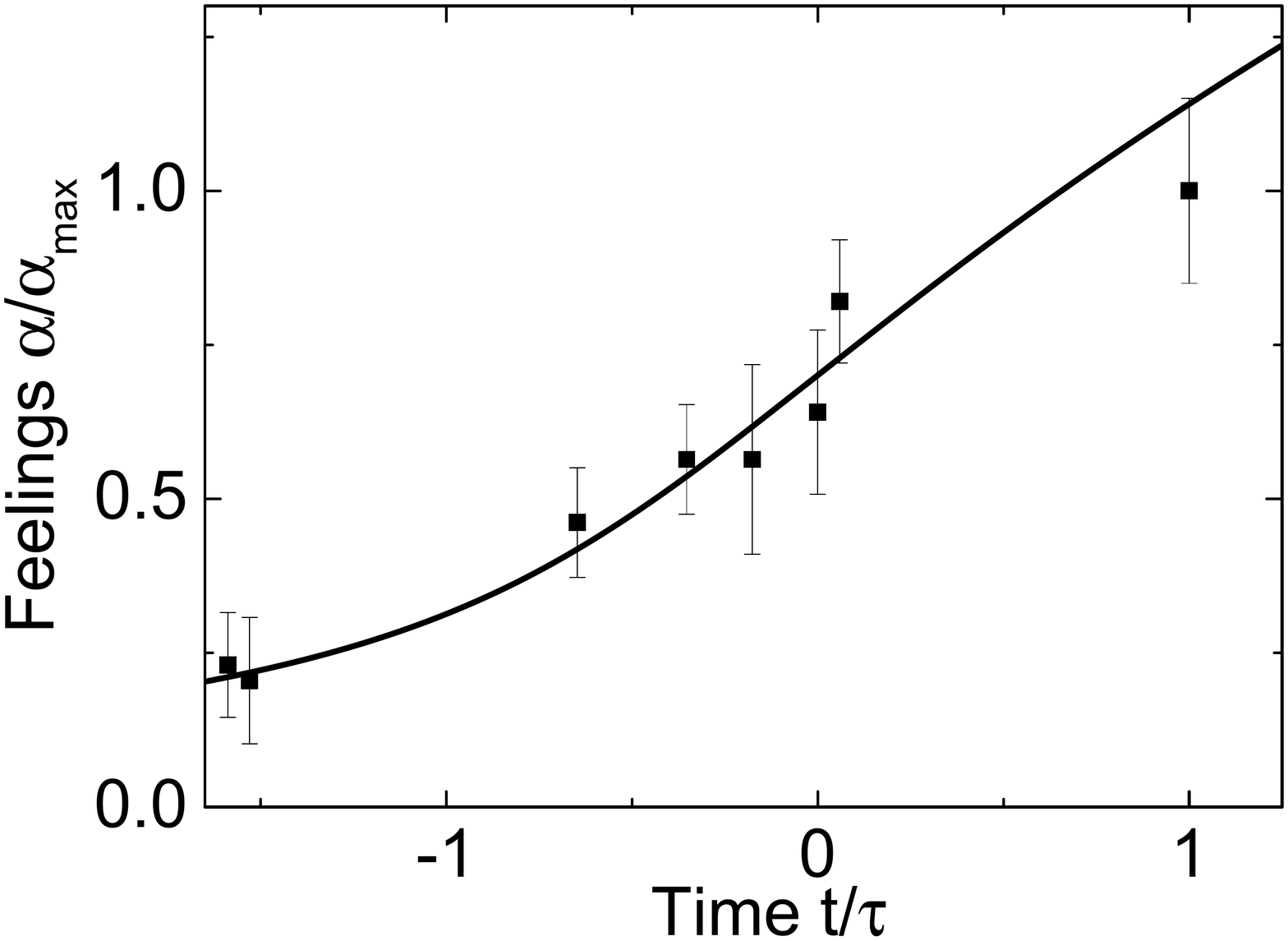}
    \caption{Love from liking. The intensity of feelings $\alpha$ for "Jane Eyre" (dots) and theory (line) with a fitting parameter $h$ (bias). Error bars indicate the standard deviation.}
    \label{fig4}
\end{figure}

\begin{figure}
    \centering
    \includegraphics[width=0.99\linewidth]{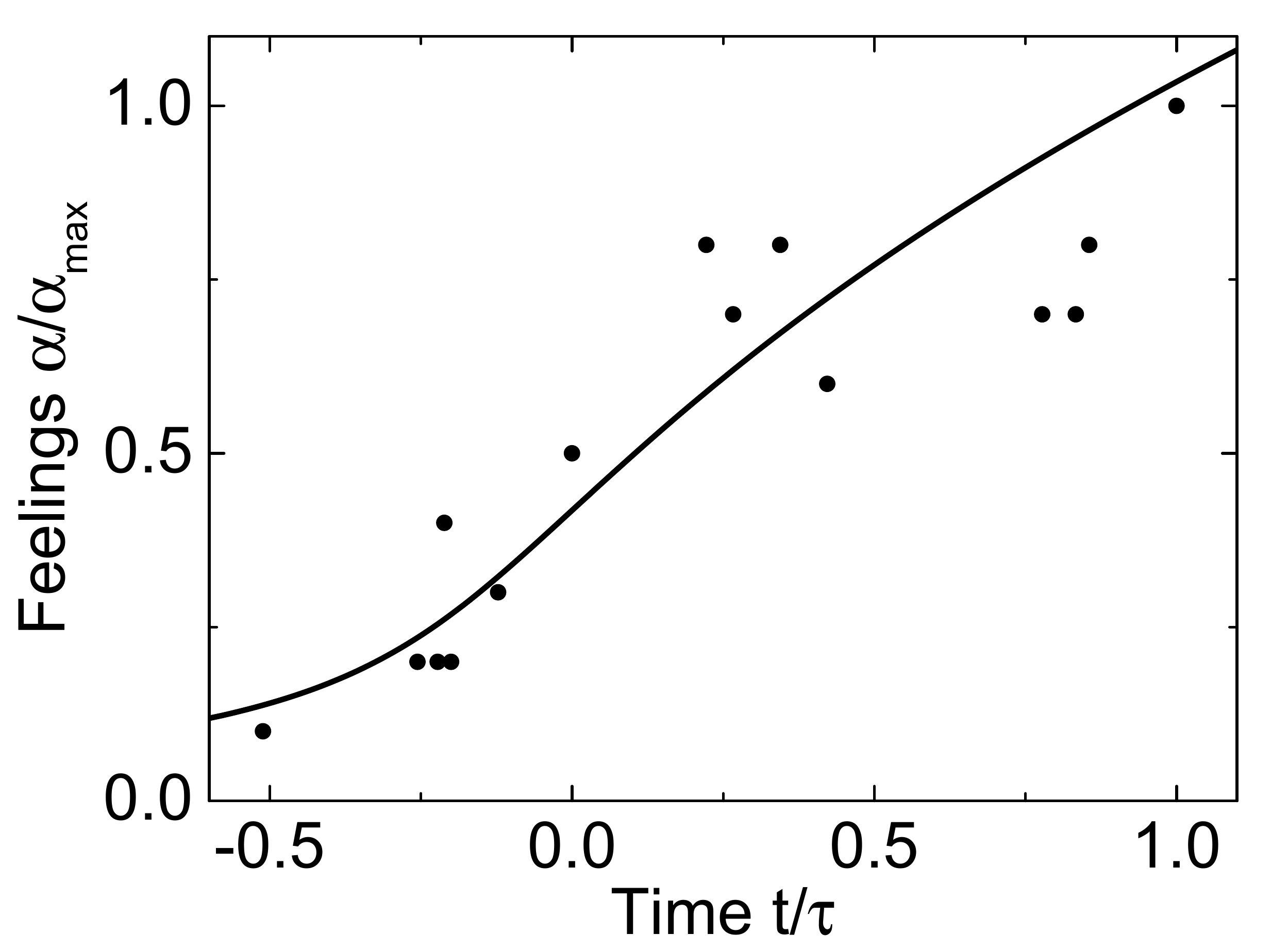}
    \caption{Love from liking. The intensity of feelings $\alpha$ for a personal diary (dots) and theory (line) with a fitting parameter $h$ (bias).}
    \label{fig5}
\end{figure}

Finally, one of the respondents has managed to find a personal diary regularly kept over about 15 years in the 1970s. She has then analyzed one example of falling in love, described in detail over several months (137 days), using the same principle: for each day, the intensity of emotions was assessed on the scale from 0 to 10. We have then rescaled the intensity from 0 to 1, as in the other examples, and taken the day where the 0.5 value has been achieved as the moment of transition, rescaling the time axis correspondingly ($\tau=90$). The results of the analysis of the diary and the fitting are presented in Fig.~\ref{fig5}. There are no error bars, since we could not ask several persons to analyze the diary because of its private nature. It is therefore difficult to estimate the uncertainty. The fitting procedure gives the bias $h\approx -0.15\pm 0.05$ (p-value $4\times 10^{-3}$). This bias is smaller than in the case of "Jane Eyre", and the consequences of this difference will be discussed below.

\section{Discussion and conclusions}

The cubic equation~\eqref{cub} can have 1 or 3 real roots, depending on the sign of the determinant 
\begin{equation}
    \Delta=-4\left(\frac{A}{2B}\right)^3-27\left(\frac{h}{4B}\right)^2
\end{equation}
($\Delta<0$ -- 1 root, $\Delta>0$ -- 3 roots). For $h=0$, the determinant changes sign with $A$, with $\Delta>0$ corresponding to the supercritical phase. The $\alpha=0$ root is unstable (maximum), but the two other roots (positive and negative) are stable. In the framework of our hypothesis it means that love ($\alpha>0$) and hate ($\alpha<0$) are both stable solutions. Once fallen in love, it is difficult to become indifferent to the beloved person; but love can be relatively easily replaced by hate. This is in agreement with another proverb, "Love And Hate Are Just One Step Apart". The difficulty to get rid of love is also confirmed by the proverb "Old love and wood will burn as soon as they get the chance." However, according to recent studies, love can disappear naturally at the timescale of several years, which is probably explained by the backwards phase transition (e.g. saturation of the brain's sensitivity to dopamine or just decrease of its level).

The situation is different for a non-zero value of $h$: while it is still possible to have 3 roots for a small value of $|h|$, the situation changes when $|h|$ is increased. For $|h|$ higher than the critical value
\begin{equation}
    h_c=\frac{2^{3/2}|A_{max}|^{3/2}}{3^{3/2}B^{1/2}}
\end{equation}
where $|A_{max}|$ is the maximal achievable value of $|A|$ for a particular person, 
the equation~\eqref{cub} always has a single root and the potential~\eqref{pot} exhibits a single minimum. It means that when love is born from a strong liking or friendship, it does not cause a strong change in the intensity of feelings at the transition, but 
it cannot turn into hate or into indifference. This theoretical prediction is illustrated in the common knowledge by the saying "The happiest marriages are between best friends". This might be an important result for applied psychology in the field of couples therapy, which only relatively recently started to focus on the importance of love \cite{willi1997significance} and on its consequences for successful relationships and marriage \cite{Vannier2017}.

For the normalization used in our work, $B=1$ and $A=(2/3)^{1/3}$, which gives $h_c=\approx 0.44$. The value found for the case of Jane Eyre $|h_{JE}|\approx 0.68$ clearly exceeds this critical value, allowing to expect that her relationship with M. Rochester should be able to withstand any hardships. This turns out indeed to be the case in the book. On the other hand, the value found for the diary $|h_{d}|\approx 0.15$ is clearly below the critical value, allowing to expect that this relationship could disappear after a certain time. This indeed turned out to be the case after 1 year. We can therefore conclude that people who do not fall in love at first sight and feel a smooth increase of their feelings, so that they might be even unsure if they are in love, should not worry too much about it: while being difficult to distinguish from friendship, their relationship is actually more secure precisely because of this.

Indeed, the transition from friendship or liking to love is so smooth that it may even go unnoticed. However, the strength of the bonds is felt immediately if one loses the object of love. In this case one understands that the feeling was actually love from the intensity of the loss that one suffers. Recent studies in psychology confirm that "love at first sight" leads to a quicker romantic involvement \cite{Barelds2007}, but the partners show more dissimilar personalities, confirming the interpretation of spontaneous symmetry breaking.

Universality implies that the same behavior should be observed for all individuals (but possibly with different scales, such as the time scale $\tau$). We stress that while the number of books that we have treated is quite small, it nevertheless provides a strong support for the hypothesis of universality, because we did not select these books. We have simply taken the ones providing a description of love at first sight and of slowly developing love that could be exploited, and the extracted behavior turned out to be in agreement with the hypothesis. We do not pretend, of course, that the present work provides a definitive proof of love being a second-order transition, or of the criticality of brain's function. More extensive studies will be necessary, both based on the analysis of texts and on the measurements of the brain activity. These studies can probably be carried out with animals, who can probably also exhibit similar phase transitions during special periods because of the same evolutionary reasons. We also note that the way how the text is perceived by the reader changes with time. For example, the comparison with torches from "Romeo and Juliet" (phrase 2) seems probably much more emotional now than when it was written. It is also interesting to note how much can be really lost in translation (see Appendix~\ref{appLily}).

To conclude, we have shown that the hypothesis that love is a second-order phase transition occurring in human brain predicts a characteristic scaling of the intensity of the feelings over time after falling in love. It is confirmed qualitatively by the common wisdom expressed in proverbs and quantitatively by the analysis of examples of "love at first sight" and "love from liking" in literature. We note a possible existence of critical "liking" or friendly attitude, above which the relations might be more stable. A more direct confirmation by the measurements of non-critical brain behavior should be possible in the future. Other possibilities include the studies of records in blogs.

\begin{acknowledgments}
We acknowledge the support of the European Union's Horizon 2020 program, through a FET Open research and innovation action under the grant agreement No. 964770 (TopoLight), project ANR Labex GaNEXT (ANR-11-LABX-0014), and of the ANR program "Investissements d'Avenir" through the IDEX-ISITE initiative 16-IDEX-0001 (CAP 20-25). 
\end{acknowledgments}

\appendix

\section{The Lily of the Valley}
\label{appLily}
The selected phrases were: 
\begin{enumerate}
    \item Aussitôt je sentis un parfum de femme qui brilla dans mon âme comme y brilla depuis la poésie orientale. 
    \item Je regardai ma voisine, et fus plus ébloui par elle que je ne l’avais été par la fête ; elle devint toute ma fête.
    \item Les plus légers détails de cette tête furent des amorces qui réveillèrent en moi des jouissances infinies
    \item tout me fit perdre l’esprit
\end{enumerate}
In this case, we had to abandon four intermediate phrases (not shown here), which seemed too direct for  modern "puritan" readers and were not qualified as manifestations of love. We have accounted for these phrases when placing the points along the time axis.

It is interesting to see how much can be lost in translation. Indeed, the English translation of phrase 2 reads: "I looked at my neighbor, and was more dazzled by that vision than I had been by the scene of the fete." The last part of the phrase ("she became all my fete") is simply absent, and without it the phrase looks much less positive and emotional. This is why we were using original phrases for our study.

\section{Jane Eyre}
\label{appendixJE}
We have selected the following paragraphs:
\begin{itemize}
    \item (Day 1) dark face, with stern features and a heavy brow; his eyes and gathered eyebrows looked ireful and thwarted just now; The new face, too, was like a new picture introduced to the gallery of memory; and it was dissimilar to all the others hanging there: firstly, because it was masculine; and, secondly, because it was dark, strong, and stern. I had it still before me when I entered Hay, and slipped the letter into the post-office; I saw it as I walked fast down-hill all the way home.
    \item (Day 2) ...decisive nose, more remarkable for character than beauty; his full nostrils, denoting, I thought, choler; his grim mouth, chin, and jaw—yes, all three were very grim, and no mistake. neither tall nor graceful.
    \item (Day 17) not quite so stern—much less gloomy; great, dark eyes, and very fine eyes, too—not without a certain change in their depths sometimes, which, if it was not softness, reminded you, at least, of that feeling. I am sure most people would have thought him an ugly man; yet there was so much unconscious pride in his port; so much ease in his demeanour; 
    \item (Day 22) I really possessed the power to amuse him, and that these evening conferences were sought as much for his pleasure as for my benefit. I had a keen delight in receiving the new ideas he offered, in imagining the new pictures he portrayed, and following him in thought through the new regions he disclosed, never startled or troubled by one noxious allusion.
    \item (Day 25) the friendly frankness, as correct as cordial, with which he treated me, drew me to him. I felt at times as if he were my relation rather than my master.  So happy, so gratified did I become with this new interest…
    \item (Day 28)  And was Mr. Rochester now ugly in my eyes? No, reader: gratitude, and many associations, all pleasurable and genial, made his face the object I best liked to see; his presence in a room was more cheering than the brightest fire. Yet I had not forgotten his faults; indeed, I could not, for he brought them frequently before me. He was proud, sardonic, harsh to inferiority of every description: in my secret soul I knew that his great kindness to me was balanced by unjust severity to many others. I believed he was naturally a man of better tendencies, higher principles, and purer tastes  I thought there were excellent materials in him; I cannot deny that I grieved for his grief, whatever that was, and would have given much to assuage it.
    \item (Day 29) Till morning dawned I was tossed on a buoyant but unquiet sea, where billows of trouble rolled under surges of joy.
    \item (Day 45) “beauty is in the eye of the gazer.” My master’s colourless, olive face, square, massive brow, broad and jetty eyebrows, deep eyes, strong features, firm, grim mouth,—all energy, decision, will,—were not beautiful, according to rule; but they were more than beautiful to me; they were full of an interest, an influence that quite mastered me,—that took my feelings from my own power and fettered them in his. I had not intended to love him; the reader knows I had wrought hard to extirpate from my soul the germs of love there detected; and now, at the first renewed view of him, they spontaneously arrived, green and strong! He made me love him without looking at me. Did I forbid myself to think of him in any other light than as a paymaster? Blasphemy against nature! Every good, true, vigorous feeling I have gathers impulsively round him. while I breathe and think, I must love him.”
\end{itemize}
We note that the days 17-25 can be attributed only approximately from the text.

\bibliography{biblio}

\end{document}